\documentclass[preprint,showpacs,preprintnumbers,amsmath,amssymb]{revtex4}

\usepackage{graphicx}
\usepackage{dcolumn}
\usepackage{bm}

\begin{document}

\preprint{APS/123-QED}

\title{Analysis of Binarized High Frequency Financial Data \\
}

\author{Naoya Sazuka}
\email{Naoya.Sazuka@jp.sony.com}
\affiliation{
Corporate Human Resources Dept, Sony Corporation,\\
4-10-18 Takanawa Minato-ku, Tokyo, 108-0074 Japan\\
}

\date{\today}

\begin{abstract}
A non-trivial probability structure is evident in the binary data extracted from the up/down price movements of very high frequency data such as tick-by-tick data for USD/JPY. In this paper, we analyze the Sony bank USD/JPY rates, ignoring the small deviations from the market price. We then show there is a similar non-trivial probability structure in the Sony bank rate, in spite of the Sony bank rate's having less frequent and larger deviations than tick-by-tick data. However, this probability structure is not found in the data which has been sampled from tick-by-tick data at the same rate as the Sony bank rate. Therefore, the method of generating the Sony bank rate from the market rate has the potential for practical use since the method retains the probability structure as the sampling frequency decreases.
\end{abstract}

\maketitle

\section{\label{sec:level1}INTRODUCTION}
In recent years, there have been various analyses of high frequency financial data which has been within the past decade become available~\cite{hf,takayasu}. High frequency financial data describe the market behavior in fine detail and they have an interesting statistical property that is not seen from the low frequency data. 

In our previous works~\cite{ohira,sazuka2}, we showed the non-trivial probability structure in the tick-by-tick data for USD/JPY with a focus on the direction of up/down price movements, as opposed to the extent of price movement. A non-trivial probability structure is not apparent from price changes themselves~\cite{sazuka}. Such a probability structure is also observed in the high frequency data of the NYSE for example GE data. In addition, we proposed the non-linear logit model that agrees well with the empirical data. 

The purpose of this paper is to understand in more detail the non-trivial probability structure of binarized high frequency data. We analyze the Sony bank USD/JPY rates, ignoring the small deviations from the market rate. We found that the non-trivial probability structure is evident in the Sony bank USD/JPY rate, even if the data have about a 20 minute interval between data points and larger price changes than tick-by-tick data.

The paper is organized as follows. In Section 2, we first explain the Sony bank and the Sony bank rate. In Section 3, we analyze the Sony bank rate. We present our conclusions in Section 4.

\section{DATA}
\subsection{Sony bank}
The Sony bank (\url{http:// moneykit.net/}), launched in 2001, is an internet-based bank for individual customers in Japan and aims to provide them with low cost services as close as possible to the professional market environment. One of their main services is online foreign-exchange trading, and this is quite popular. As of July 2005, the Sony bank deals with eight currencies and they have about 130,000 foreign-currency accounts, which is about 30$\%$ of the total number of accounts. On the other hand, foreign-currency accounts are only about 3$\%$ of all accounts in conventional banks.

There are three main reasons for the popularity of the bank's foreign-exchange trading. Firstly, trades can be made on the web 24 hours a day, 7 day a week. In conventional banks in Japan, however, customers actually have to go to a branch and can trade only during weekday business hours. Secondly, customers can trade at the Sony bank's rate which at all times closely reflects the market rate. In conventional banks, the trading rate for individual customers is fixed at one time in the morning for the day's trading. Finally, the bank's low transaction cost. For example, the trading cost of the Sony bank for USD/JPY is 0.25 yen per dollar, which is about a quarter of the trading cost of conventional banks

\subsection{Sony bank rate}

\begin{figure}[ht]
\begin{center}
\includegraphics [scale=.6]{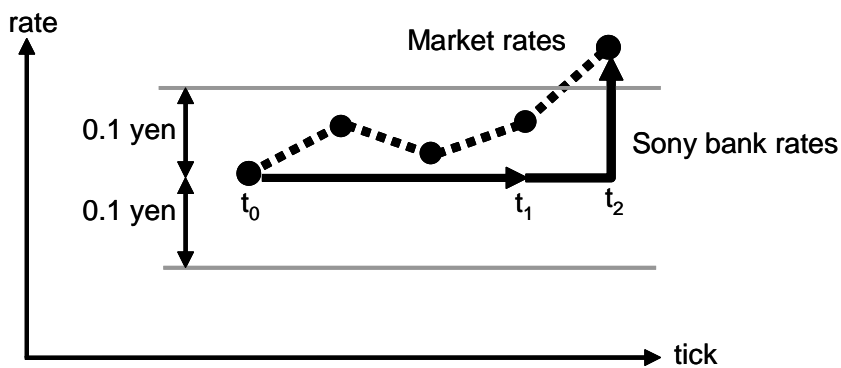}
\caption{Update of the Sony bank rate for USD/JPY. The Sony bank USD/JPY rate is a solid line and the market rate is a broken line. In this figure, the Sony bank USD/JPY rate is fixed from t0 to t1 and is updated to the market rate at t2, as the market rate exceeds the range of 0.1 yen at t2 based on the market rate at t0.}
\label{fig:1}
\end{center}
\end{figure}

In this paper, we analyze the Sony bank rate for USD/JPY. The Sony bank rate is the foreign-currency exchange rate that customers having a foreign-currency account of the Sony bank can deal at. The Sony bank rate is dependent on the market rate but is independent of the customers' orders. If the USD/JPY market rate changes more than 0.1 yen, the Sony bank rate is updated to the market rate. Figure~\ref{fig:1} is a schematic of the updating of the Sony bank rate for USD/JPY. If the market rate changes along the broken line, the Sony bank rate follows the solid line. Suppose the market rate (broken line) is the same as the Sony bank rate (solid line) at the time t0. The Sony bank rate stays flat from t0 to t1, as the market rate is in the range of 0.1 yen based on the market rate at t0. When the market rate exceeds the range of 0.1 yen at t2, the Sony bank rate is updated to the market rate.

Since the Sony bank rate ignores small deviations in the market price by applying its updating rule, the Sony bank rate has less frequent and larger price changes than tick-by-tick data. The comparison is shown in Table~\ref{tab:table1s}. The smallest price change, which is the majority of the price changes, is 0.1 yen for the Sony bank rate. On the other hand, 40$\%$ of price changes in tick-by-tick data are zero deviations between data points. We ignore such cases in order to focus on only the direction of up/down price movements. Then tick-by-tick data moves by a 0.01 yen unit.

\begin{table}[ht]
\begin{center}
\caption{The Sony bank USD/JPY rate and tick-by-tick data for USD/JPY}
\label{tab:table1s}
{
\begin{tabular}{|l|c|c|} \hline
& Sony bank rate & tick-by-tick data \\ \hline
Number of data a day & $\sim$70 & $\sim$10,000 \\ \hline
The smallest price change & 0.1 yen & 0.01 yen \\ \hline
Average interval between data & $\sim$20 minutes & $\sim$7 seconds \\ \hline
\end{tabular}
}
\end{center}
\end{table}

\section{Analysis}
\begin{figure}[ht]
\begin{center}
\includegraphics [scale=.8]{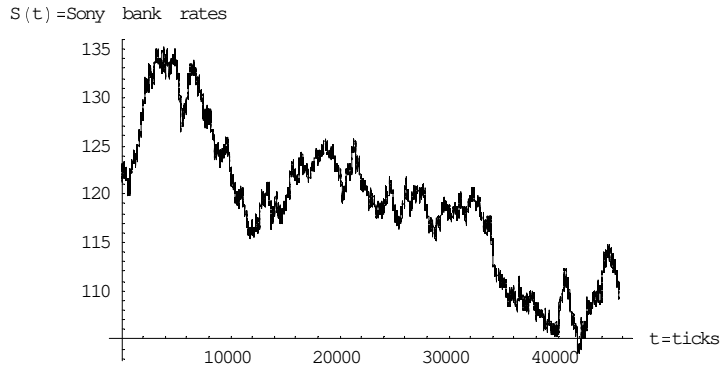}
\caption{The Sony bank USD/JPY rates ploted against ticks for the period 22/10/2001 to 31/05/2005.}
\label{fig:2}
\end{center}
\end{figure}

We analyze the Sony bank USD/JPY rates for the period of 22/10/2001 to 31/05/2005 (Figure~\ref{fig:2}). The data is composed of value $S(t)$ of yen value per dollar at the Sony bank tick step $t$. The number of data is 45,577.

As analyzed in our previous works~\cite{ohira,sazuka2}, we first binarize data extracting the direction of the up/down price movement in the following way: $X(t)=+1$ if $S(t+1)-S(t)>0$ and $X(t)=-1$ if $S(t+1)-S(t)<0$. According to the update rule, the Sony bank rate do not have zero deviations between data points $(S(t+1)-S(t)=0)$ not like tick-by-tick data. Next we compute the conditional probabilities of binary data $X(t)$ upto 4th order.

\begin{figure}[ht]
\begin{center}
\includegraphics [scale=.6]{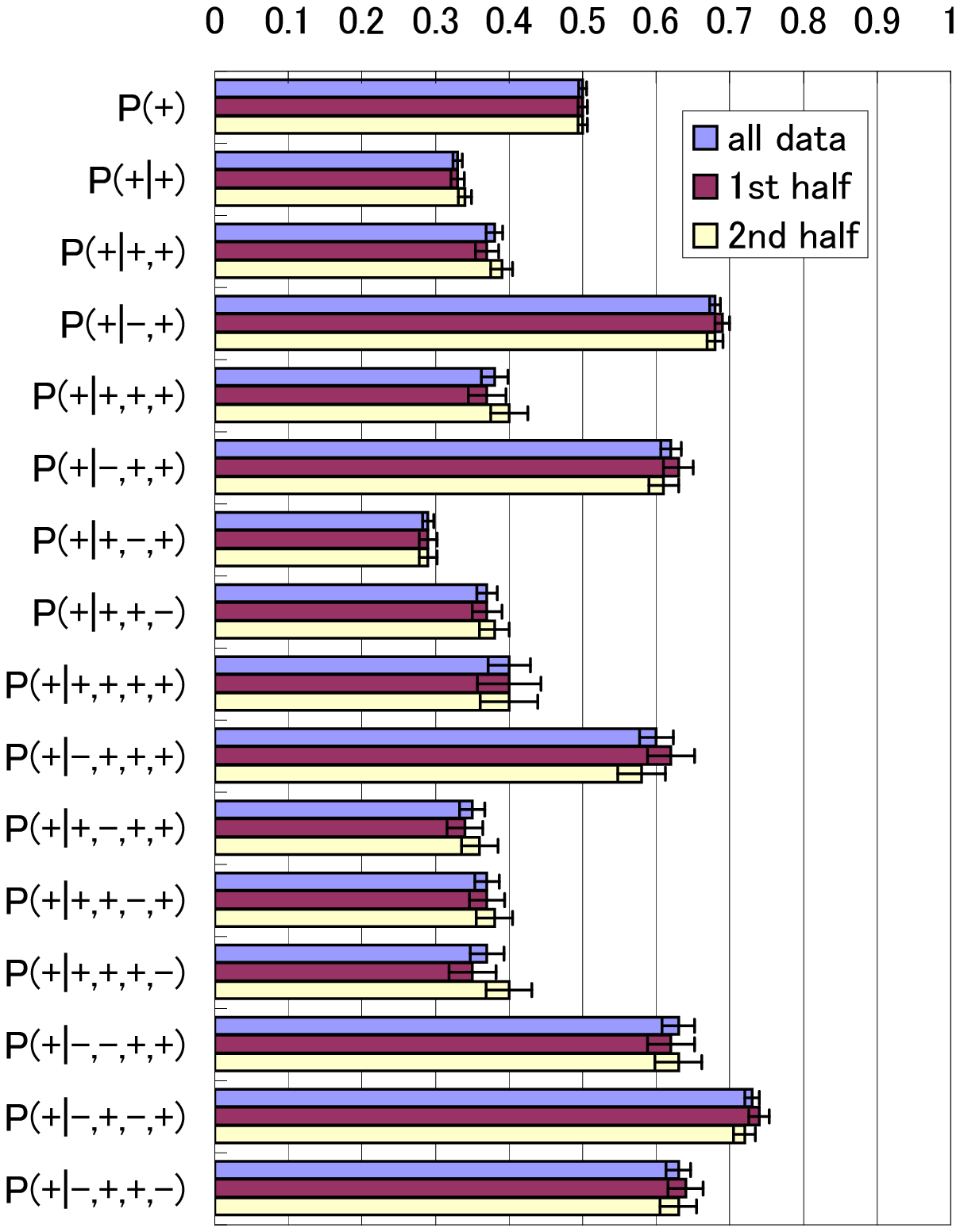}
\caption{Probability structure of the Sony bank USD/JPY rate. Each conditional probability has three bars. The upper one is the result of whole period 22/10/2001 to 31/05/2005. The middle is for the first half of the data. The lower is for the second half of the data. The error bars represent a 95$\%$ confidence interval on the mean}
\label{fig:3}
\end{center}
\end{figure}

Then we show the non-trivial conditional probability structure of the binary data $X(t)$. Figure~\ref{fig:3} is the probability structure of the Sony bank rates for whole period, and the first half of the data and the second half of the data. The data was divided into two periods to compare the tendencies. We show only ``+" side of the conditional probabilities because of the up/down symmetry. Although it would seem obvious that if the market is rising, there must be more +'s than -'s, in fact, if we only have the data which is triggers and change in the Sony bank rate, there will be over the long run a almost equal number of + and - changes. This non-trivial probability structure is not only in the first order conditional probabilities but also higher orders. Especially, it is likely to happen a regular ``zigzag" motion of alternating up and down, for example $P(+|-,+,-,+)$ is 73$\%$, rather than irregular ``zigzag" motions, for example $P(+|-,+,-,+)$ is 63$\%$ and a motion in one direction, for example $P(+|+,+,+,+)$ is 40$\%$.

It is interesting that the Sony bank rate has a probability structure, considering that the average interval between data points is about 20 minutes. However, this probability structure is not found in the data which has been sampled from tick-by-tick data at the same rate as the Sony bank rate. Zhang looked similar higher order statistics and found that the probability structure was not observed in the data recoded every 30 minutes~\cite{zhang}. In addition, the probability structure of the Sony bank rate is not the effect of the negative first-order auto correlation in the second scale tick-by-tick data, since many transactions occur during 20 minutes. Therefore, the method of generating the Sony bank rate from the market rate can retain the probability structure as a sampling frequency decreases. 

The probability structure of the Sony bank rate is not only based on less frequent sampling but also larger deviations than tick-by-tick data. As shown in an Table~\ref{tab:table1s}, the probability structure of tick-by-tick data is mostly 0.01 yen price changes within 7 seconds on an average and sometimes even shorter. This may be too frequent and small for practical use. However, the Sony bank rate essentially reflects mostly 0.1 yen price changes and about 20 minutes interval between price chages. Therefore, there is a better chance to take advantage of the Sony bank rate.

In addition, we compare the probability structure of GE stock data on NYSE shown in~\cite{sazuka2}. The data was collected in December 2002. The number of GE data for one day is about 180,000, which is more frequent than the tick-by-tick data for USD/JPY.  Figure 4 shows that the reliability of the probability structure increases as the sampling frequency increases, comparing tick-by-tick data for USD/JPY, the Sony bank USD/JPY rate and GE data on NYSE. The exceptions are regular ``zigzag" motions such as $P(+|-,+),P(+|+,-,+),P(+|-,+,-,+)$. In these cases, the probability structure of the Sony bank rate is more reliable than the tick-by-tick USD/JPY rate.

\begin{figure}[ht]
\begin{center}
\includegraphics[scale=.6]{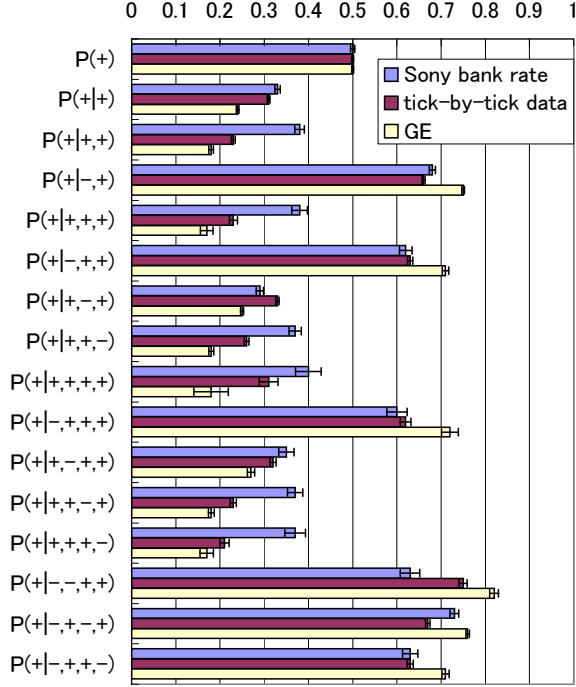}
\caption{
Comparison of the bias among the tick-by-tick data for GE on NYSE, tick-by-tick data for USD/JPY and Sony bank USD/JPY rates. The reliability of the probability structure increaces as the frequency of the data increaces. The error bars represent a $95\%$ confidence interval on the mean.}
\label{fig:4}
\end{center}
\end{figure}

\section{CONCLUSION}
We analyzed the Sony bank USD/JPY rate, which closely reflects the market rate, with a focus on the up/down price movement, as opposed to the extent of price movement. The Sony bank has a lower sampling rate and shows larger price change deviations than the market rate. The main result is that a probability structure is observed in the Sony bank rate, in spite of the interval between data points of the Sony bank rate is about 20 minutes. However, this probability structure is not found in the data which has been sampled from tick-by-tick data at the same rate as the Sony bank rate. Therefore, the method of generating the Sony bank rate from the market rate can retain the probability structure as the sampling frequency decreases. There is a better chance to take advantage of the Sony bank rate in practical use.

\begin{acknowledgments}
I would like to thank Shigeru Ishi, President of the Sony bank, for kindly providing the Sony bank data. The NYSE data are from King's College London.
\end{acknowledgments}

\bibliography{apssamp}

\end{document}